\documentclass[a4paper,10pt]{article}
\usepackage{amssymb}
\usepackage{graphicx}
\usepackage{amsmath}

 \DeclareFontFamily{U}{rsf}{}
\DeclareFontShape{U}{rsf}{m}{n}{
  <5> <6> rsfs5 <7> <8> <9> rsfs7 <10-> rsfs10}{}
\DeclareMathAlphabet\Scr{U}{rsf}{m}{n} \makeatletter
\@addtoreset{equation}{section} \makeatother

\def\be{\begin{equation}}
\def\ee{\end{equation}}
\def\ba{\begin{array}}
\def\ea{\end{array}}

\newcommand{\bea}{\begin{eqnarray}}
\newcommand{\eea}{\end{eqnarray}}

\begin{document}

\begin{titlepage}
 \thispagestyle{empty}
\begin{flushright}
     \hfill{CERN-PH-TH/2011-028}\\
 \end{flushright}

 \vspace{100pt}

 \begin{center}
     { \huge{\bf      {Freudenthal Duality and\\\vspace{10pt}Generalized Special Geometry }}}

     \vspace{25pt}

     {\Large {Sergio Ferrara$^{a,b}$, Alessio Marrani$^{a}$ and Armen Yeranyan$^{b,c}$}}

     \vspace{40pt}

  {\it ${}^a$ Physics Department, Theory Unit, CERN,\\
     CH -1211, Geneva 23, Switzerland;\\
     \texttt{sergio.ferrara@cern.ch}\\
     \texttt{Alessio.Marrani@cern.ch}}

     \vspace{10pt}

    {\it ${}^b$ INFN - Laboratori Nazionali di Frascati,\\
     Via Enrico Fermi 40, I-00044 Frascati, Italy\\
     \texttt{ayeran@lnf.infn.it}}

     \vspace{10pt}



{\it ${}^c$ Department of Physics, Yerevan State University\\
Alex Manoogian St. 1, Yerevan, 0025, Armenia}\\

     \vspace{10pt}

     \vspace{60pt}

     {ABSTRACT}

 \vspace{10pt}
 \end{center}
Freudenthal duality, introduced in \cite{Duff-Freud-1} and defined
as an anti-involution on the dyonic charge vector in $d=4$
space-time dimensions for those dualities admitting a quartic
invariant, is proved to be a symmetry not only of the classical
Bekenstein-Hawking entropy but also of the critical points of the
black hole potential.

Furthermore, Freudenthal duality is extended to any generalized
special geometry, thus encompassing all $\mathcal{N}>2$
supergravities, as well as $\mathcal{N}=2$ generic special geometry,
not necessarily having a coset space structure.
\end{titlepage}

\section{\label{Intro}Introduction}

\textit{Freudenthal duality} was introduced by Borsten, Dahanayake, Duff and
Rubens in \cite{Duff-Freud-1} as an anti-involutive operator on the
representation space of the dyonic charge $\mathcal{Q}^{A}$, defined as%
\footnote{%
For $\mathcal{I}_{4}<0$, the definition (\ref{1}) is actually the opposite
of definition given in \cite{Duff-Freud-1}; however, this is immaterial for
all subsequent treatment, because $\phi _{H}\left( -\mathcal{Q}\right) =\phi
_{H}\left( \mathcal{Q}\right) $ and (\ref{5}) holds.}
\begin{equation}
\widehat{\mathcal{Q}}^{A}\left( \mathcal{Q}\right) \equiv \mathbb{C}^{AB}%
\frac{\partial \sqrt{\left| \mathcal{I}_{4}\left( \mathcal{Q}\right) \right|
}}{\partial \mathcal{Q}^{B}},  \label{1}
\end{equation}
where $\mathcal{I}_{4}$ is the invariant polynomial quartic in $\mathcal{Q}$%
, and $\mathbb{C}$ is the $2n\times 2n$ symplectic metric ($\mathbb{C}^{T}=-%
\mathbb{C}$, $\mathbb{C}^{2}=-\mathbb{I}$, $n$ denoting the number of vector
fields of the theory)
\begin{equation}
\mathbb{C}\equiv \left(
\begin{array}{cc}
0 & -\mathbb{I} \\
\mathbb{I} & 0
\end{array}
\right) .
\end{equation}
Note that $\widehat{\mathcal{Q}}$ defined by (\ref{1}) is homogeneous of
degree one in $\mathcal{Q}$.

The basic properties of such an operator are \cite{Duff-Freud-1}
\begin{eqnarray}
\widehat{\widehat{\mathcal{Q}}} &=&-\mathcal{Q~}\text{(\textit{%
anti-involution})};  \label{2} \\
\mathcal{I}_{4}\left( \widehat{\mathcal{Q}}\right) &=&\mathcal{I}_{4}\left(
\mathcal{Q}\right) .  \label{3}
\end{eqnarray}
Note that this duality holds for both signs of $\mathcal{I}_{4}$, thus
embracing both BPS and non-BPS (``large'') BHs. On the other hand, the
definition (\ref{1}) is ill-defined when $\mathcal{I}_{4}=0$, corresponding
to the so-called ``small'' BHs, whose Bekenstein-Hawking \cite{BH} entropy $%
S $ (and thus area of the horizon) vanishes. Also, in the present paper we
do not consider the issue of charge quantization and the related
restrictions for the implementation of Freudenthal duality \cite
{Duff-Freud-1} (see also remark 1 in Sec. \ref{Conclusion}).

In mathematical literature, groups with a symplectic representation $\mathbf{%
R}$ admitting a completely symmetric rank-$4$ invariant structure $\mathbf{q}
$ such that the invariant polynomial $\mathcal{I}_{4}$ can be defined as%
\footnote{%
The normalization of $\mathbf{q}$ used here is the same one adopted in \cite
{Duff-Freud-1}, and it thus differs by a factor $2$ with respect to the one
adopted \textit{e.g.} in \cite{Brown-E7}, \cite{Exc-Reds} and \cite{Irred-1}.%
}
\begin{equation}
\mathcal{I}_{4}\left( \mathcal{Q}\right) \equiv \frac{1}{2}\left. \mathbf{q}%
\left( \mathcal{Q}_{1},\mathcal{Q}_{2},\mathcal{Q}_{3},\mathcal{Q}%
_{4}\right) \right| _{\mathcal{Q}_{1}=\mathcal{Q}_{2}=\mathcal{Q}_{3}=%
\mathcal{Q}_{4}\equiv \mathcal{Q}},  \label{I4}
\end{equation}
are sometimes called \textit{``groups of type }$E_{7}$\textit{''} (see
\textit{e.g.} \cite{Brown-E7}; in the case of $E_{7}$, $\mathbf{R}$ is the
fundamental representation $\mathbf{56}$). \textit{``Groups of type }$E_{7}$%
\textit{'' } are \textit{at least} all the $U$-duality groups of $\mathcal{N}%
=2$, $d=4$ supergravity theories with symmetric vector multiplets' scalar
manifold, as well as of four-dimensional $\mathcal{N}>2$-extended
supergravities (all based on symmetric spaces). These include theories
(namely, \textit{minimally coupled} $\mathcal{N}=2$ \cite{Luciani}, and $%
\mathcal{N}=3$ \cite{N=3} supergravity) in which the basic invariant
polynomial is quadratic: $\left| \mathcal{I}_{2}\right| =\sqrt{\left|
\mathcal{I}_{4}\right| }$, and thus it should replace $\mathcal{I}_{4}$
itself in Eqs. (\ref{1}) and (\ref{3}).

In the $\mathcal{N}=2$ case, the properties (\ref{2})-(\ref{3}) are a
consequence of the symplectic structure of \textit{special K\"{a}hler
geometry} endowing vector multiplets' scalar manifold (see \textit{e.g.}
\cite{CDF-rev}, and Refs. therein), while for $\mathcal{N}>2$ theories they
can be traced back to what one may call \textit{generalized special geometry}%
. This corresponds to the sigma models which can be consistently coupled to
vector fields, as discussed in \cite{GZ} in general, and then treated in
\cite{ADF,FK-N=8} in the context of extended supergravities.

The main result of the present paper is the finding that the Freudenthal
duality is in fact a property which can be defined with a dependence on the
scalar fields $\phi $, parametrizing the whole scalar manifold $\mathbf{M}$,
\textit{in any generalized special geometry}. Let us recall the definition
of the (positive-definite) black hole potential \cite{FGK}
\begin{equation}
V\left( \phi ,\mathcal{Q}\right) \equiv -\frac{1}{2}\mathcal{Q}^{T}\mathcal{M%
}\left( \phi \right) \mathcal{Q},  \label{V}
\end{equation}
where $\mathcal{M}$ is the symmetric, real, negative-definite $2n\times 2n$
matrix made of the vector couplings \cite{BGM,CDF-rev,FGK}, which depends on
the scalar fields $\phi $ and it is symplectic:
\begin{equation}
\mathcal{M}\mathbb{C}\mathcal{M}=\mathbb{C}.  \label{gen-id}
\end{equation}
In the treatment given below, we will show that $V$ and $\mathcal{M}$ are
Freudenthal invariant\footnote{%
Throughout our treatment, the subscript ``$H$'' denotes evaluation at the
event horizon of the extremal black hole solution under consideration.} at
the attractor points of $V$ (determining the fixed, attractor \cite
{AM-Refs,FGK} values of the scalar fields):
\begin{eqnarray}
\left. V^{2}\right| _{\partial V=0}\left( \phi _{H}\left( \mathcal{Q}\right)
,\mathcal{Q}\right) &=&\left| \mathcal{I}_{4}\left( \mathcal{Q}\right)
\right| =\left| \mathcal{I}_{4}\left( \widehat{\mathcal{Q}}\right) \right| ;
\label{4} \\
\mathcal{M}_{H}\left( \widehat{\mathcal{Q}}\right) &=&\mathcal{M}_{H}\left(
\mathcal{Q}\right) ,  \label{5-bis}
\end{eqnarray}
where $\mathcal{M}_{H}\left( \mathcal{Q}\right) \equiv $ $\mathcal{M}\left(
\phi _{H}\left( \mathcal{Q}\right) \right) $.

The results (\ref{4}) and (\ref{5-bis}) will be generalized to a scalar
field dependent framework in Sec. \ref{1-Off-Shell}. Indeed, an
interpolating (scalar field dependent) generalization of the Freudenthal
duality will be defined in Sec. \ref{1-Off-Shell} as a symmetry of the
effective $1$-dimensional action obtained as reduction of the bosonic sector
of the action of an \textit{ungauged} $\mathcal{N}\geqslant 2$, $d=4$
Maxwell-Einstein supergravity by exploiting the space-time symmetries
(staticity, spherical symmetry) of the asymptotically flat extremal dyonic
black hole background \cite{BGM,FGK}.

Remarkably, this reasoning can be extended to non-symmetric $\mathcal{N}=2$
special geometries. Within this broad class of theories, definition (\ref{1}%
) and Eq. (\ref{4}) enjoy the following generalization\footnote{%
Our definition of $S\left( \mathcal{Q}\right) $ is as follows:
\begin{equation*}
S\left( \mathcal{Q}\right) \equiv V\left( \phi _{H}\left( \mathcal{Q}\right)
\right) \equiv V_{H}\left( \mathcal{Q}\right) =\frac{1}{4\pi }A_{H},
\end{equation*}
where in the last step we used the Bekenstein-Hawking entropy-area formula
\cite{BH} ($A_{H}$ denotes the area of the unique event horizon of the
extremal BH).}:
\begin{eqnarray}
\widehat{\mathcal{Q}}^{A} &\equiv &\mathbb{C}^{AB}\frac{\partial S\left(
\mathcal{Q}\right) }{\partial \mathcal{Q}^{B}};  \label{6} \\
S\left( \mathcal{Q}\right) &\equiv &\left. V\right| _{\partial V=0}\left(
\phi _{H}\left( \mathcal{Q}\right) ,\mathcal{Q}\right) .  \label{7}
\end{eqnarray}
$S$ is the classical Bekenstein-Hawking black hole entropy \cite{BH}, which
is homogeneous of degree two in the charges:
\begin{equation}
\mathcal{Q}^{A}\frac{\partial S\left( \mathcal{Q}\right) }{\partial \mathcal{%
Q}^{A}}=2S\left( \mathcal{Q}\right) .  \label{8}
\end{equation}
$\widehat{\mathcal{Q}}$ defined by (\ref{6}) is the \textit{symplectic
gradient} of $S$. Therefore, (\ref{7}) is the statement that $S$ is
invariant when replacing $\mathcal{Q}$ with the symplectic gradient of $S$.
Note that in general, $S$ can be a rather complicated function of $\mathcal{Q%
}$ (often not exactly computable, as well), whereas for groups of type $%
E_{7} $ \cite{Brown-E7}, $S\left( \mathcal{Q}\right) =\sqrt{\left| \mathcal{I%
}_{4}\left( \mathcal{Q}\right) \right| }$.

The generalization of Eqs. (\ref{3}) and (\ref{4}) yields, by means of
definition (\ref{6}), the result
\begin{equation}
S\left( \mathcal{Q}\right) =S\left( \mathbb{C}\frac{\partial S}{\partial
\mathcal{Q}}\right) .  \label{9}
\end{equation}
This result is general; for instance, it can explicitly be checked to hold
in $\mathcal{N}=2$, $d=4$ not symmetric nor homogeneous few-moduli models,
such as the ones considered in \cite{Shmakova}.

These results are consequences of the symplectic structure of (generalized)
special geometry, characterized by the fundamental identity \cite{FK-N=8}
\begin{equation}
\mathcal{MV}_{a}=i\mathbb{C}\mathcal{V}_{a},  \label{10}
\end{equation}
where $\mathcal{V}_{a}\left( \phi \right) $ is the $2n\times 1$ vector of
complex symplectic sections, related to $\mathcal{M}$ through the identity%
\footnote{%
For a specialization to the $\mathcal{N}=8$ case, see \textit{e.g.}\ \cite
{FK-N=8,CFGn-1}.} ($a=1,...,n$) \cite{FK-N=8}
\begin{equation}
\mathcal{M}+i\mathbb{C}=2\left( \mathbb{C}\mathcal{V}_{a}\right) \left(
\overline{\mathcal{V}}^{aT}\mathbb{C}\right) .  \label{Thu-1}
\end{equation}
In the $\mathcal{N}=2$ case, this identity can be rewritten as
\begin{equation}
\mathcal{M}+i\mathbb{C}=2\left( \mathbb{C}\mathbf{V}\right) \left( \overline{%
\mathbf{V}}^{T}\mathbb{C}\right) +2g^{i\overline{j}}\left( \mathbb{C}%
\overline{\mathbf{V}}_{\overline{j}}\right) \left( \mathbf{V}_{i}^{T}\mathbb{%
C}\right) ,  \label{SKG-id}
\end{equation}
where $\mathcal{V}_{a}\equiv \left( \mathbf{V},\overline{\mathbf{V}}_{%
\overline{i}}\right) $, and $\mathbf{V}\equiv e^{\mathcal{K}/2}\left(
X^{\Lambda },F_{\Lambda }\right) ^{T}$ is the $2n\times 1$ vector of
K\"{a}hler-covariantly holomorphic sections ($\mathcal{K}$ denotes the
K\"{a}hler potential, $\mathbf{V}_{i}\equiv D_{i}\mathbf{V}\equiv \left(
\partial _{i}+\frac{1}{2}\partial _{i}\mathcal{K}\right) \mathbf{V}$, and $%
i=1,....,n_{V}=n-1$, with $n_{V}$ standing for the number of vector
multiplets).\medskip

The paper is organized as follows.

In Sec. \ref{1-On-Shell} we consider Freudenthal duality, and we show that,
as asserted above, it does not hold only for symmetric scalar manifolds (for
which a quartic polynomial $\mathcal{I}_{4}$ can be introduced), but rather
it can be defined for \textit{all} generalized special geometries, and for
all non-degenerate critical points of the effective black hole potential $V$%
, namely for those critical points, BPS or not, for which the supporting
charge orbit is ``large'', \textit{i.e.} $S\left( \mathcal{Q}\right) \neq 0$%
. The case already known in literature \cite{Duff-Freud-1} is then recovered
in Sec. \ref{Duff-Groups-of-Type-E7}.

As an example, we consider $\mathcal{N}=2$ BPS attractors, for which the
sections $\mathbf{V}_{BPS}$ are given by a complexified charge vector, whose
real and imaginary parts are related by Freudenthal duality. This can be
traced back to the existence of an \textit{interpolating} (scalar field
dependent) Freudenthal duality, which is introduced in Sec. \ref{1-Off-Shell}%
, in which the results of Sec. \ref{1-On-Shell} are extended to the whole
scalar manifold $\mathbf{M}$ pertaining to \textit{generalized special
geometry}. The effective black hole potential $V$ (despite being
non-invariant under the Freudenthal duality treated in Sec. \ref{1-On-Shell}%
) is shown to be invariant under the \textit{interpolating} Freudenthal
duality. It is worth here remarking that the evaluation of the \textit{%
interpolating} Freudenthal duality at the critical points of $V$ themselves
consistently yields the Freudenthal duality treated in\ Sec. \ref{1-On-Shell}
(which in turn, for groups of type $E_{7}$, reduces to the one originally
introduced in \cite{Duff-Freud-1}).

Given two (static, asymptotically flat, spherically symmetric, dyonic)
extremal black hole solutions whose dyonic charge vectors are related by
Freudenthal duality, they are characterized by the same critical points of $%
V $, and thus they have the same classical Bekenstein-Hawking entropy. This
holds despite the fact that their (attractor) scalar flows are different,
and that $V$ is not invariant under the scalar field independent Freudenthal
duality discussed in Sec. \ref{1-On-Shell}.

Sec. \ref{Conclusion} contains some remarks, in particular concerning the
quantization of dyonic vector $\mathcal{Q}$, the generalization of the
quartic invariant $\mathcal{I}_{4}$ to non-symmetric special K\"{a}hler
geometries, the relation to quaternionic K\"{a}hler geometry and its $%
\mathcal{N}=2$, $d=4$ harmonic superspace formulation. \textit{\ }

\section{\label{1-On-Shell}Freudenthal Duality}

We start by considering the black hole (BH) effective potential \cite{FGK}
\begin{equation}
V\equiv -\frac{1}{2}\mathcal{Q}^{T}\mathcal{MQ},  \label{8-bis}
\end{equation}
whose criticality conditions read
\begin{equation}
\left( \partial _{i}V\right) _{H}=\left( D_{i}V\right) _{H}=0,~\forall
i\Leftrightarrow \mathcal{Q}^{T}\left( D_{i}\mathcal{M}\right) _{H}\mathcal{Q%
}=0,~\forall i.  \label{11}
\end{equation}
It then follows that
\begin{equation}
\left. \frac{\partial V}{\partial \mathcal{Q}}\right| _{H}=-\mathcal{M}_{H}%
\mathcal{Q}=-\mathcal{M}_{H}\mathcal{Q}-\frac{1}{2}\mathcal{Q}^{T}\left(
D_{i}\mathcal{M}\right) _{H}\mathcal{Q}\frac{\partial \phi _{H}^{i}\left(
\mathcal{Q}\right) }{\partial \mathcal{Q}}=\frac{\partial V_{H}\left(
\mathcal{Q}\right) }{\partial \mathcal{Q}}=\frac{\partial S\left( \mathcal{Q}%
\right) }{\partial \mathcal{Q}},  \label{13}
\end{equation}
where $V_{H}\left( \mathcal{Q}\right) \equiv V\left( \phi _{H}\left(
\mathcal{Q}\right) ,\mathcal{Q}\right) $, $\mathcal{M}_{H}\left( \mathcal{Q}%
\right) \equiv \mathcal{M}\left( \phi _{H}\left( \mathcal{Q}\right) \right) $%
, and the same for $\left( D_{i}\mathcal{M}\right) _{H}$.

As anticipated in the Introduction, the symplectic structure of $\mathcal{N}%
\geqslant 2$-extended supergravity theories in $d=4$ space-time dimensions
\cite{CDF-rev,FK-N=8} allows one to introduce a duality operator $\frak{F}_{%
\mathcal{Q}}$, dubbed (after \cite{Duff-Freud-1}) \textit{Freudenthal duality%
}, acting on $Sp\left( 2n,\mathbb{R}\right) $ vectors. Its action on the
charge vector $\mathcal{Q}$ is defined as
\begin{equation}
\frak{F}_{\mathcal{Q}}:\mathcal{Q}\longmapsto \widehat{\mathcal{Q}}%
_{H}\equiv \mathbb{C}\frac{\partial S\left( \mathcal{Q}\right) }{\partial
\mathcal{Q}}=-\mathbb{C}\mathcal{M}\left( \phi _{H}\left( \mathcal{Q}\right)
\right) \mathcal{Q}.  \label{14}
\end{equation}
As a consequence of the result (\ref{5-bis}) and of the symplectic nature of
$\mathcal{M}$ (\ref{2}), the action of $\frak{F}_{\mathcal{Q}}$ on $\mathcal{%
Q}$ is \textit{anti-involutive}:
\begin{eqnarray}
\frak{F}_{\mathcal{Q}}^{2} &\equiv &\frak{F}_{\widehat{\mathcal{Q}}%
_{H}\left( \mathcal{Q}\right) }:\mathcal{Q}\longmapsto \widehat{\widehat{%
\mathcal{Q}}}\left( \mathcal{Q}\right) \equiv \widehat{\mathcal{Q}}%
_{H}\left( \widehat{\mathcal{Q}}_{H}\left( \mathcal{Q}\right) \right)  \notag
\\
&=&\mathbb{C}\mathcal{M}\left( \phi _{H}\left( \widehat{\mathcal{Q}}%
_{H}\right) \right) \mathbb{C}\mathcal{M}\left( \phi _{H}\left( \mathcal{Q}%
\right) \right) \mathcal{Q}=\mathbb{C}\mathcal{M}\left( \phi _{H}\left(
\mathcal{Q}\right) \right) \mathbb{C}\mathcal{M}\left( \phi _{H}\left(
\mathcal{Q}\right) \right) \mathcal{Q}  \notag \\
&=&-\mathcal{Q}.  \label{15}
\end{eqnarray}
Note that $\frak{F}_{\mathcal{Q}}$ acts trivially on the symplectic sections
$\mathcal{V}_{a,H}$ at critical points of $V$:
\begin{equation}
\frak{F}_{\mathcal{Q}}:\mathcal{V}_{a,H}\longmapsto \mathcal{V}_{a,H}.
\label{17}
\end{equation}

We anticipate that the critical points of the effective BH potential $%
V\left( \phi ,\mathcal{Q}\right) $ (defining the attractor configurations of
scalar fields at the BH horizon) are also $\frak{F}_{\mathcal{Q}}$%
-invariant, as it will be proved at the end of Sec. \ref{1-Off-Shell}. An
illustrative example is provided by the BPS attractor Eqs. in $\mathcal{N}=2$%
, $d=4$ special geometry, corresponding to the local vanishing of all
\textit{matter charges}
\begin{equation}
\left( D_{i}Z\right) _{H}\equiv Z_{i,H}=0~\forall i,  \label{BPS-large-cond}
\end{equation}
where $Z\equiv \mathcal{Q}^{T}\mathbb{C}\mathbf{V}$ is the complex $\mathcal{%
N}=2$ central charge. By means of (\ref{SKG-id}), an equivalent algebraic
reformulation of (\ref{BPS-large-cond}) reads \cite{FK-1,D-1,FGimK}
\begin{equation}
2i\left( \overline{Z}\mathbf{V}\right) _{BPS}=\mathcal{Q}+i\mathbb{C}%
\mathcal{M}_{H}\mathcal{Q}=\mathcal{Q}-i\widehat{\mathcal{Q}}_{H}.
\label{18}
\end{equation}
Let's now $\frak{F}_{\mathcal{Q}}$-transform this equation. By defining the $%
1\times \left( n_{V}+1\right) $ vector
\begin{equation}
\mathcal{Z}_{a}\equiv \mathcal{Q}^{T}\mathbb{C}\mathcal{V}_{a}\equiv \left(
Z,Z_{i}\right) ,
\end{equation}
and using the general identity (\ref{10}), one can show that
\begin{equation}
\frak{F}_{\mathcal{Q}}:\mathcal{Z}_{a,H}\longmapsto \widehat{\mathcal{Z}}%
_{a,H}\equiv \widehat{\mathcal{Q}}_{H}^{T}\mathbb{C}\mathcal{V}_{a}=-i%
\mathcal{Z}_{a,H}.  \label{18-bis}
\end{equation}
By recalling the anti-involutivity property (\ref{15}), $\mathbf{V}_{H,BPS}$
defined by Eq. (\ref{18}) then turns out to be invariant under $\frak{F}_{%
\mathcal{Q}}$, consistent with (\ref{17}). Therefore, BPS attractor Eqs. (%
\ref{18}) (and consequently their solutions, corresponding to the
non-degenerate BPS critical points of $V$), are $\frak{F}_{\mathcal{Q}}$%
-invariant. By using (\ref{18-bis}), this can also be checked for the
equivalent differential conditions (\ref{BPS-large-cond}), of course.

As evident from the treatment given below, the criticality conditions of $V$
(in turn splitting under supersymmetry in the three $\mathcal{N}=2$ classes
of BPS, non-BPS $Z_{H}=0$ and non-BPS $Z_{H}\neq 0$ attractor Eqs.; see
\textit{e.g.} \cite{BFGM1}) are generally $\frak{F}_{\mathcal{Q}}$%
-invariant, and this result in general holds when generalizing $\frak{F}_{%
\mathcal{Q}}$ to $\frak{F}_{\phi }$ (see Sec. \ref{1-Off-Shell}), and for in
all (generalized) special geometries.

A remarkable consequence of (\ref{14}) and (\ref{15}) is that \textit{the
Bekenstein-Hawking BH entropy }$S$\textit{\ is }$\frak{F}_{\mathcal{Q}}$%
\textit{-invariant}. Indeed, (\ref{14}) implies that
\begin{gather}
\widehat{\mathcal{Q}}_{H}^{T}\mathbb{C}\mathcal{Q}=\mathcal{Q}\frac{\partial
S}{\partial \mathcal{Q}}=-\mathcal{Q}^{T}\mathcal{M}_{H}\mathcal{Q}%
=2V_{H}\left( \mathcal{Q}\right) =2S\left( \mathcal{Q}\right) ;  \label{19}
\\
\Updownarrow  \notag \\
S\left( \mathcal{Q}\right) =\frac{1}{2}\left\langle \widehat{\mathcal{Q}}%
_{H},\mathcal{Q}\right\rangle \Rightarrow \widehat{S}\left( \mathcal{Q}%
\right) \equiv S\left( \widehat{\mathcal{Q}}_{H}\right) =S\left( \mathcal{Q}%
\right) .  \label{20}
\end{gather}
Note that (\ref{19}) is consistent with the fact that $S\left( \mathcal{Q}%
\right) =V\left( \phi _{H}\left( \mathcal{Q}\right) ,\mathcal{Q}\right)
\equiv V_{H}\left( \mathcal{Q}\right) $ is homogeneous of degree two (and
thus even) in the charges $\mathcal{Q}$.

\textit{(\ref{20}) is a general result, which holds in any (}$\mathcal{N}%
\geqslant 2$\textit{, }$d=4$\textit{) generalized special geometry. }Within
this broad class of theories, any two BH solutions whose dyonic charge
vectors are related by $\frak{F}_{\mathcal{Q}}$ have the same critical
points of $V$ (as it will be proved at the end of Sec. \ref{1-Off-Shell})
and the same classical entropy $S$. This holds despite the fact that $%
V\left( \phi ,\mathcal{Q}\right) \neq V\left( \phi ,\widehat{\mathcal{Q}}%
_{H}\right) $ (on the other hand, Eqs. (\ref{19}) and (\ref{20}) imply that $%
V_{H}\left( \mathcal{Q}\right) \neq V_{H}\left( \widehat{\mathcal{Q}}%
_{H}\right) $). It is worth here noting that the action of $\frak{F}_{%
\mathcal{Q}}$ on rank-$4$ classical charge orbits preserves both the
supersymmetry properties and the rank \footnote{%
Consistent with \cite{Duff-Freud-1}, here by \textit{rank} we mean the rank
of any representative $\mathcal{Q}$ of the orbit as element of the
corresponding Freudenthal triple system, as defined in \cite{rank-FTS}.} of
these orbits.

\subsection{\label{Duff-Groups-of-Type-E7}The case of Groups of Type $E_{7}$}

For those generalized special geometries \cite{FK-N=8} related to groups of
type $E_{7}$ \cite{Brown-E7}, which are endowed with an invariant structure $%
\mathbf{q}$ defining the quartic invariant $\mathcal{I}_{4}\mathcal{\ }$(\ref
{I4}), $\frak{F}_{\mathcal{Q}}$ reduces to the \textit{non-polynomial}
Freudenthal duality introduced in \cite{Duff-Freud-1}, here denoted by $%
\frak{F}_{\mathcal{Q},\mathbf{q}}$ (recall definition (\ref{1}) and Footnote
1):
\begin{equation}
\frak{F}_{\mathcal{Q},\mathbf{q}}:\mathcal{Q}\longmapsto \widehat{\mathcal{Q}%
}_{H,\mathbf{q}}\left( \mathcal{Q}\right) \equiv -\mathbb{C}\mathcal{M}_{H,%
\mathbf{q}}\mathcal{Q}=\mathbb{C}\frac{\partial \sqrt{\left| \mathcal{I}%
_{4}\right| }}{\partial \mathcal{Q}}=\epsilon \frac{1}{2\sqrt{\left|
\mathcal{I}_{4}\right| }}\frac{\partial \mathcal{I}_{4}}{\partial \mathcal{Q}%
},  \label{21}
\end{equation}
where $\mathcal{I}_{4}\equiv \epsilon \left| \mathcal{I}_{4}\right| $.

From points \textit{a)}, \textit{e)}, \textit{d)} and \textit{g)} of Lemma
11 of \cite{Brown-E7}, various results on all possible evaluations of $%
\mathbf{q}$ on $\mathcal{Q}$ and $\widehat{\mathcal{Q}}_{H,\mathbf{q}}$
follow:
\begin{eqnarray}
\mathbf{q}\left( \mathcal{Q}^{3}\widehat{\mathcal{Q}}_{H,\mathbf{q}}\right)
&=&0=\mathbf{q}\left( \mathcal{Q}\widehat{\mathcal{Q}}_{H,\mathbf{q}%
}^{3}\right) ;  \label{21-2} \\
\mathbf{q}\left( \mathcal{Q}^{2}\widehat{\mathcal{Q}}_{H,\mathbf{q}%
}^{2}\right) &=&\frac{1}{3}\mathbf{q}\left( \mathcal{Q}^{4}\right) ,
\label{21-4}
\end{eqnarray}
and
\begin{equation}
\mathbf{q}\left( \widehat{\mathcal{Q}}_{H,\mathbf{q}}^{4}\right) =\mathbf{q}%
\left( \mathcal{Q}^{4}\right) \overset{(\ref{I4})}{\Leftrightarrow }\frak{F}%
_{\mathcal{Q},\mathbf{q}}\left( \mathcal{I}_{4}\left( \mathcal{Q}^{4}\right)
\right) \equiv \mathcal{I}_{4}\left( \widehat{\mathcal{Q}}_{H,\mathbf{q}%
}^{4}\right) =\mathcal{I}_{4}\left( \mathcal{Q}^{4}\right) .  \label{21-5}
\end{equation}
(\ref{21-5}) consistently expresses the $\frak{F}_{\mathcal{Q},\mathbf{q}}$%
-invariance of the $U$-invariant polynomial $\mathcal{I}_{4}$ defined by (%
\ref{I4}).

\section{\label{1-Off-Shell}Interpolating (Field-Dependent) Formulation}

As a consequence of the basic identities of generalized special geometry
\cite{FK-N=8}, one can introduce an alternative operator $\frak{F}_{\phi }$,
which is \textit{scalar-dependent} and it is defined \textit{all over the
scalar manifold} $\mathbf{M}$. The action of $\frak{F}_{\phi }$ on the
charge vector $\mathcal{Q}$ and on the symplectic sections $\mathcal{V}_{a}$
is respectively defined as
\begin{eqnarray}
\frak{F}_{\phi } &:&\mathcal{Q}\longmapsto \widehat{\mathcal{Q}}_{\phi
}\left( \phi \right) \equiv \mathbb{C}\frac{\partial V\left( \phi ,\mathcal{Q%
}\right) }{\partial \mathcal{Q}}=-\mathbb{C}\mathcal{M}\left( \phi \right)
\mathcal{Q};  \label{30} \\
\frak{F}_{\phi } &:&\mathcal{V}_{a}\longmapsto \mathcal{V}_{a}.  \label{32}
\end{eqnarray}
Note that (\ref{32}) implies that the scalar fields $\phi $, coordinatizing $%
\mathbf{M}$, are trivially $\frak{F}_{\phi }$-invariant.

As a consequence of the symplectic nature (\ref{2}) of $\mathcal{M}\left(
\phi \right) $, it is immediate to see that the non-trivial action of $\frak{%
F}_{\phi }$ on $\mathcal{Q}$ is \textit{anti-involutive}:
\begin{equation}
\frak{F}_{\phi }^{2}:\mathcal{Q}\longmapsto \widehat{\widehat{\mathcal{Q}}}%
_{\phi }\left( \phi \right) =-\mathcal{Q}.  \label{31}
\end{equation}

Note that (\ref{14}) and (\ref{17}) are nothing but (\ref{30}) and (\ref{32}%
) evaluated at the critical points of $V$, respectively. Furthermore, the
general ientity (\ref{Thu-1}) and the trivial action (\ref{32}) of $\frak{F}%
_{\phi }$ on the symplectic sections $\mathcal{V}_{a}$ imply the matrix $%
\mathcal{M}\left( \phi \right) $ to be $\frak{F}_{\phi }$-invariant:
\begin{equation}
\widehat{\mathcal{M}}_{\phi }\left( \phi \right) \equiv \mathcal{M}\left(
\frak{F}_{\phi }\left( \mathcal{V}_{a}\left( \phi \right) \right) \right) =%
\mathcal{M}\left( \phi \right) .  \label{Thu-2}
\end{equation}
At the critical points of $V$, Eq. (\ref{Thu-2}) reduces to Eq. (\ref{5-bis}%
).

As anticipated, a consequence of (\ref{30}) and (\ref{31}), which can
ultimately be traced back to the basic properties of generalized special
geometry, is that \textit{the effective BH potential }$V\left( \phi ,%
\mathcal{Q}\right) $\textit{\ is }$\frak{F}_{\phi }$\textit{-invariant}.
Indeed, (\ref{30}) implies that
\begin{gather}
\widehat{\mathcal{Q}}_{\phi }^{T}\mathbb{C}\mathcal{Q}=-\mathcal{Q}^{T}%
\mathcal{M}\left( \phi \right) \mathcal{Q}=2V\left( \phi ,\mathcal{Q}\right)
;  \label{35} \\
\Updownarrow   \notag \\
V\left( \phi ,\mathcal{Q}\right) =\frac{1}{2}\left\langle \widehat{\mathcal{Q%
}}_{\phi },\mathcal{Q}\right\rangle ;  \label{35-bis} \\
\Downarrow   \notag \\
\widehat{V}_{\phi }\left( \phi ,\mathcal{Q}\right) \equiv V\left( \phi ,%
\widehat{\mathcal{Q}}_{\phi }\right) =\frac{1}{2}\widehat{\widehat{\mathcal{Q%
}}}_{\phi }^{T}\mathbb{C}\widehat{\mathcal{Q}}_{\phi }=\frac{1}{2}\widehat{%
\mathcal{Q}}_{\phi }^{T}\mathbb{C}\mathcal{Q}=V\left( \phi ,\mathcal{Q}%
\right) .  \label{36}
\end{gather}
Note that (\ref{35}) is consistent with the fact that $V$ is a polynomial
homogeneous of degree two (and thus even) in the charges $\mathcal{Q}$. Note
that (\ref{5}) and (\ref{36}) immediately imply that
\begin{equation}
V_{H}\left( \mathcal{Q}\right) \equiv V_{H}\left( \phi _{H}\left( \mathcal{Q}%
\right) ,\mathcal{Q}\right) =V_{H}\left( \phi _{H}\left( \widehat{\mathcal{Q}%
}_{H}\right) ,\mathcal{Q}\right) \equiv V_{H}\left( \widehat{\mathcal{Q}}%
_{H}\left( \mathcal{Q}\right) \right) =\frac{1}{2}\left\langle \widehat{%
\mathcal{Q}}_{H}\left( \mathcal{Q}\right) ,\mathcal{Q}\right\rangle .
\end{equation}

\textit{(\ref{36}) is a general result, which holds in any (}$\mathcal{N}%
\geqslant 2$\textit{, }$d=4$\textit{) symplectic geometry.\smallskip }

Let us now prove that the critical points of $V\left( \phi ,\mathcal{Q}%
\right) $ coincide with the critical points of $V\left( \phi ,\widehat{%
\mathcal{Q}}_{H}\left( \mathcal{Q}\right) \equiv \frak{F}_{\mathcal{Q}%
}\left( \mathcal{Q}\right) \right) $, namely that the critical points $\phi
_{H}\left( \mathcal{Q}\right) $ of $V\left( \phi ,\mathcal{Q}\right) $ are $%
\frak{F}_{\mathcal{Q}}$-invariant.

First, we observe that by differentiating the general identity (\ref{gen-id}%
) with respect to the scalar fields $\phi $ and using (\ref{gen-id}) again,
one obtains
\begin{equation}
\frac{\partial \mathcal{M}\left( \phi \right) }{\partial \phi }=\mathbb{%
\mathcal{M}\left( \phi \right) \mathbb{C\frac{\partial \mathcal{M}\left(
\phi \right) }{\partial \phi }C}\mathcal{M}\left( \phi \right) },
\end{equation}
which multiplied by $-\frac{1}{2}\mathcal{Q}^{T}$ on the left and by $%
\mathcal{Q}$ on the right yields, by recalling definitions (\ref{V}) and (%
\ref{30}):
\begin{equation}
\frac{\partial V\left( \phi ,\mathcal{Q}\right) }{\partial \phi }=-\frac{1}{2%
}\mathcal{Q}^{T}\frac{\partial \mathcal{M}\left( \phi \right) }{\partial
\phi }\mathcal{Q}=\frac{1}{2}\widehat{\mathcal{Q}}_{\phi }^{T}\mathbb{%
\mathbb{\frac{\partial \mathcal{M}\left( \phi \right) }{\partial \phi }}}%
\widehat{\mathcal{Q}}_{\phi }.  \label{n-1}
\end{equation}
Therefore, the results (\ref{36}) and (\ref{n-1}) imply
\begin{equation}
\frac{\partial V\left( \phi ,\mathcal{Q}\right) }{\partial \phi }=\frac{%
\partial V\left( \phi ,\widehat{\mathcal{Q}}_{\phi }\right) }{\partial \phi }%
,  \label{n-2}
\end{equation}
whose evaluation at $\phi _{H}\left( \mathcal{Q}\right) $ yields by
definition:
\begin{equation}
0=\left. \frac{\partial V\left( \phi ,\mathcal{Q}\right) }{\partial \phi }%
\right| _{\phi _{H}\left( \mathcal{Q}\right) }=\frac{1}{2}\widehat{\mathcal{Q%
}}_{\phi _{H}\left( \mathcal{Q}\right) }^{T}\left. \mathbb{\mathbb{\frac{%
\partial \mathcal{M}\left( \phi \right) }{\partial \phi }}}\right| _{\phi
_{H}\left( \mathcal{Q}\right) }\widehat{\mathcal{Q}}_{\phi _{H}\left(
\mathcal{Q}\right) }\equiv \frac{1}{2}\widehat{\mathcal{Q}}_{H}^{T}\left.
\mathbb{\mathbb{\frac{\partial \mathcal{M}\left( \phi \right) }{\partial
\phi }}}\right| _{\phi _{H}\left( \mathcal{Q}\right) }\widehat{\mathcal{Q}}%
_{H}.  \label{n-3}
\end{equation}
As a final step, let us consider
\begin{equation}
V\left( \phi ,\widehat{\mathcal{Q}}_{H}\right) \equiv -\frac{1}{2}\widehat{%
\mathcal{Q}}_{H}^{T}\mathcal{M}\left( \phi \right) \widehat{\mathcal{Q}}_{H},
\label{n-4}
\end{equation}
whose derivative with respect to scalar fields reads, evaluated at $\phi
_{H}\left( \widehat{\mathcal{Q}}_{H}\right) \equiv \widehat{\phi }_{H}\left(
\mathcal{Q}\right) $ reads by definition
\begin{equation}
0\equiv \left. \frac{\partial V\left( \phi ,\widehat{\mathcal{Q}}_{H}\right)
}{\partial \phi }\right| _{\phi _{H}\left( \widehat{\mathcal{Q}}_{H}\right)
}=-\frac{1}{2}\widehat{\mathcal{Q}}_{H}^{T}\left. \frac{\partial \mathcal{M}%
\left( \phi \right) }{\partial \phi }\right| _{\phi _{H}\left( \widehat{%
\mathcal{Q}}_{H}\right) }\widehat{\mathcal{Q}}_{H}.  \label{n-5}
\end{equation}
The comparison of (\ref{n-3}) and (\ref{n-5}) implies that both $\phi
_{H}\left( \mathcal{Q}\right) $ and $\phi _{H}\left( \widehat{\mathcal{Q}}%
_{H}\right) $ are critical points of $V$. On the other hand, in all $%
\mathcal{N}\geqslant 2$, $d=4$ extended supergravities (vector multiplets')
scalar fields can be parametrised projectively in terms of symplectic
sections $\mathcal{V}_{a}$. Thus, the $\frak{F}_{\mathcal{Q}}$-invariance of
$\mathcal{V}_{a,H}$ expressed by (\ref{17}) implies the $\frak{F}_{\mathcal{Q%
}}$-invariance of the critical points $\phi _{H}\left( \mathcal{Q}\right) $
of the effective BH potential $V\left( \phi ,\mathcal{Q}\right) $:
\begin{equation}
\frak{F}_{\mathcal{Q}}\left( \phi _{H}\left( \mathcal{Q}\right) \right)
\equiv \phi _{H}\left( \frak{F}_{\mathcal{Q}}\left( \mathcal{Q}\right)
\right) =\phi _{H}\left( \mathcal{Q}\right) ~~\blacksquare  \label{5}
\end{equation}
\smallskip

Thus, as anticipated, two $\frak{F}_{\mathcal{Q}}$\textit{-dual} BH
solutions (namely, two BH solutions whose dyonic charge vectors are related
by the scalar field-independent Freudenthal duality $\frak{F}_{\mathcal{Q}}$%
) have different \textit{off-shell} effective BH potentials (but with the
same critical points) and same entropy $S$, despite the fact their
(attractor) scalar flows are generally different.\medskip

As mentioned above, it is immediate to check that all definitions and
results concerning $\frak{F}_{\phi }$ consistently reduce, along the
geometrical \textit{locus} in $\mathbf{M}$ defined by the (non-degenerate)
critical points of $V$, to the analogous definitions and results for $\frak{F%
}_{\mathcal{Q}}$ considered in Sec. \ref{1-On-Shell}. Namely:
\begin{equation}
\left. \frak{F}_{\phi }\right| _{\partial V\left( \phi ,\mathcal{Q}\right)
=0,V\neq 0}\equiv \frak{F}_{\phi _{H}\left( \mathcal{Q}\right) }=\frak{F}_{%
\mathcal{Q}},
\end{equation}
and thus:
\begin{equation}
\widehat{\mathcal{Q}}_{H}\left( \mathcal{Q}\right) =\widehat{\mathcal{Q}}%
_{\phi _{H}\left( \mathcal{Q}\right) }.
\end{equation}

\section{\label{Conclusion}Final Remarks}

\textbf{1. On charge quantization}

The algebraic nature of the $\mathcal{N}=2$ BPS Attractor Eqs. (\ref{18})
sheds light on the relevance of Freudenthal duality in presence of
Dirac-Schwinger-Zwanzinger dyonic charge quantization conditions, namely in
the case in which $\mathcal{Q}$ is integer. Indeed, by requiring (along the
lines of the analysis of \cite{Duff-Freud-1}) $\widehat{\mathcal{Q}}_{H}$ to
be an integer as well, it follows that (within a suitably K\"{a}hler
gauge-fixing in which $2i\overline{Z}\equiv 1$) $\mathbf{V}_{BPS}$ is a
\textit{complex integer} (namely, $\mathbf{V}_{BPS}\in \mathbb{Z}+i\mathbb{Z}
$).\medskip

\textbf{2. On the generalization of }$\mathcal{I}_{4}$

The rank-$4$ completely symmetric invariant structure $\mathbf{q}$ (\ref{I4}%
) characterizing all ``groups of type $E_{7}$'' \cite{Brown-E7} (whose
suitable real forms are the $U$-duality groups of $d=4$ supergravity
theories with symmetric scalar manifolds) has a generalization to the case
of arbitrary $\mathcal{N}=2$ \textit{non-symmetric} vector multiplets'
special K\"{a}hler scalar manifolds $\mathbf{M}$. This is obtained by
considering the function $\mathcal{I}_{4,\mathcal{N}=2,symm}$ (given by Eq.
(5.36) of \cite{CFMZ1}):
\begin{equation}
\mathcal{I}_{4,\mathcal{N}=2,symm}\left( \phi ,\mathcal{Q}\right) =\left(
\left| Z\right| ^{2}-Z_{i}\overline{Z}^{i}\right) ^{2}+\frac{2}{3}i\left(
ZN_{3}\left( \overline{Z}\right) -\overline{Z}\overline{N}_{3}\left(
Z\right) \right) -g^{i\overline{i}}C_{ijk}\overline{C}_{\overline{i}%
\overline{l}\overline{m}}\overline{Z}^{j}\overline{Z}^{k}Z^{\overline{l}}Z^{%
\overline{m}},  \label{I4-N=2-symm}
\end{equation}
where $N_{3}\left( \overline{Z}\right) \equiv C_{ijk}\overline{Z}^{i}%
\overline{Z}^{j}\overline{Z}^{k}$, with $C_{ijk}$ denoting the $C$-tensor of
special K\"{a}hler geometry. $\mathcal{I}_{4,\mathcal{N}=2,symm}$ is an
homogeneous polynomial in $\mathcal{Q}$; note also that (\ref{I4-N=2-symm})
is independent on the choice of the symplectic frame and also manifestly
invariant under diffeomorphisms in $\mathbf{M}$.

As stated in \cite{CFMZ1}, $\mathcal{I}_{4,\mathcal{N}=2,symm}$ is
independent on $\phi $ \textit{at least} for symmetric $\mathbf{M}$, whereas
generally it is $\phi $-dependent; its fourth derivative with respect to
charges defines the following rank-$4$ completely symmetric ($\mathcal{Q}$%
-independent) tensor
\begin{equation}
\Omega _{ABCD}\left( \phi \right) \equiv \frac{1}{12}\frac{\partial ^{4}%
\mathcal{I}_{4,\mathcal{N}=2,symm}\left( \phi ,\mathcal{Q}\right) }{\partial
\mathcal{Q}^{A}\partial \mathcal{Q}^{B}\partial \mathcal{Q}^{C}\partial
\mathcal{Q}^{D}}\Leftrightarrow \mathcal{I}_{4,\mathcal{N}=2,symm}\left(
\phi ,\mathcal{Q}\right) =\frac{1}{2}\Omega _{ABCD}\left( \phi \right)
\mathcal{Q}^{A}\mathcal{Q}^{B}\mathcal{Q}^{C}\mathcal{Q}^{D}.  \label{Omega}
\end{equation}
For \textit{symmetric} special K\"{a}hler manifolds $\Omega _{ABCD}=\mathbf{q%
}_{ABCD}$ defined by (\ref{I4}) is an invariant structure of the charge
representation $\mathcal{Q}$ of ``groups of type $E_{7}$'' \cite{Brown-E7},
and thus $\mathcal{I}_{4,\mathcal{N}=2,symm}=\mathcal{I}_{4}\left( \mathcal{Q%
}\right) $.

It is here worth recalling that the general relation between $\mathcal{I}_{4,%
\mathcal{N}=2,symm}\left( \phi ,\mathcal{Q}\right) $ (\ref{I4-N=2-symm}) and
the square of the effective BH potential (\ref{V}) \cite{FGK}
\begin{equation}
V^{2}\left( \phi ,\mathcal{Q}\right) =\frac{1}{4}\mathcal{M}_{(AB}\left(
\phi \right) \mathcal{M}_{CD)}\left( \phi \right) \mathcal{Q}^{A}\mathcal{Q}%
^{B}\mathcal{Q}^{C}\mathcal{Q}^{D}=\left( \left| Z\right| ^{2}+Z_{i}%
\overline{Z}^{i}\right) ^{2}
\end{equation}
enjoys the general expression at the various classes of (non-degenerate)
critical points of $V$ itself in a \textit{generic} $\mathcal{N}=2$ special
K\"{a}hler geometry:
\begin{equation}
\left. \mathcal{I}_{4,\mathcal{N}=2,symm}\right| _{H}=V_{H}^{2}-\frac{32}{3}%
\left| Z\right| _{H}^{2}\left( Z_{i}\overline{Z}^{i}\right) _{H}.
\label{crit-V-gen-rel}
\end{equation}
Namely:

\begin{itemize}
\item  at BPS critical points, defined by (\ref{BPS-large-cond}), it holds
that \cite{CFMZ1}
\begin{equation}
\left. \mathcal{I}_{4,\mathcal{N}=2,symm}\right| _{H}=\left| Z\right|
_{H}^{4}=V_{H}^{2};
\end{equation}

\item  at those non-BPS critical points defined by $Z_{H}=0$, it holds that
\cite{CFMZ1}
\begin{equation}
\left. \mathcal{I}_{4,\mathcal{N}=2,symm}\right| _{H}=\left( Z_{i}\overline{Z%
}^{i}\right) _{H}^{2}=\left[ g^{i\overline{j}}\left( \partial _{i}Z\right)
\overline{\partial }_{\overline{j}}\overline{Z}\right] _{H}^{2}=V_{H}^{2};
\end{equation}

\item  at the general class of non-BPS critical points with $Z_{H}\neq 0$,
the relation (\ref{crit-V-gen-rel}) holds, simplifying to
\begin{equation}
\left. \mathcal{I}_{4,\mathcal{N}=2,symm}\right| _{H}=-16\left| Z\right|
_{H}^{4}=-V_{H}^{2}  \label{Delta=0}
\end{equation}
when $\left( Z_{i}\overline{Z}^{i}\right) _{H}=3\left| Z\right| _{H}^{2}$,
as \textit{e.g.} it holds for \textit{symmetric} special K\"{a}hler
spaces.\medskip
\end{itemize}

\textbf{3. On the relation to special quaternionic geometry}

As noticed in \cite{CFMZ1}, $\mathcal{I}_{4,\mathcal{N}=2,symm}$ is
remarkably related to the \textit{geodesic potential} defined in the $%
d=4\rightarrow 3$ dimensional reduction of the considered $\mathcal{N}=2$
theory. Under such a reduction, the $d=4$ vector multiplets' special
K\"{a}hler manifold $\mathbf{M}$ (dim$_{\mathbb{C}}=n_{V}$) enlarges to a
\textit{special} quaternionic K\"{a}hler manifold $\frak{M}$ (dim$_{\mathbb{H%
}}=n_{V}+1$) given by $c$-map \cite{CFG,Ferrara-Sabharwal} of $\mathbf{M}$
itself : $\frak{M}=c\left( \mathbf{M}\right) $. By specifying Eq. (5.36) of
\cite{CFMZ1} in the \textit{special coordinates'} symplectic frame, $%
\mathcal{I}_{4,\mathcal{N}=2,symm}\left( \phi ,\mathcal{Q}\right) $ matches
the opposite of the function $h$ defined by Eq. (4.42) of \cite{dWVVP},
within the analysis of \textit{special} quaternionic K\"{a}hler geometry.

This relation can be strengthened by observing that the tensor $\Omega
_{ABCD}$ defined by (\ref{Omega}) is proportional to the $\Omega $-tensor of
quaternionic geometry, related to the Riemann tensor by Eq. (15) of \cite
{Bagger-Witten}.\medskip

\textbf{4. On the relation to }$\mathcal{N}=2$\textit{,} $d=4$\textbf{\
harmonic superspace}

For \textit{symmetric} $\mathbf{M}$, it holds that
\begin{equation}
\mathcal{I}_{4,\mathcal{N}=2,symm}=\mathcal{I}_{4}\left( \mathcal{Q}\right) =%
\frac{1}{2}\mathbf{q}\left( \mathcal{Q}^{4}\right) =\mathcal{L}^{+4},
\label{HSS-rel}
\end{equation}
where $\mathcal{L}^{+4}$ is the \textit{quaternionic potential} of symmetric
quaternionic K\"{a}hler non-compact spaces \cite{Wolf,A}, determined in \cite
{HSS-1,HSS-2} (see also \cite{G-HSS}) with $\mathcal{N}=2$, $d=4$ harmonic
superspace techniques, in which the $d=4$ BH dyonic charge vector $\mathcal{Q%
}$ becomes the $SU\left( 2\right) $ harmonic coordinate $Q^{+}$. Note that
for the $c$-map of the non-compact $\mathbb{CP}^{n}$ special K\"{a}hler
manifold, given by the special quaternionic K\"{a}hler manifold
\begin{equation}
c\left( \mathbb{CP}^{n}\right) =\frac{SU\left( 2,n+1\right) }{SU\left(
2\right) \times SU\left( n+1\right) \times U\left( 1\right) },
\end{equation}
$\mathcal{L}^{+4}$ becomes a perfect square (see Eq. (11.5) of \cite{HSS-2}%
), consistent with the observation below Eq. (\ref{3}). When considering
\textit{special} quaternionic K\"{a}hler spaces (\textit{i.e.} the images of
special K\"{a}hler spaces through $c$-map) \cite{CFG, Ferrara-Sabharwal},
the coordinates $\mathcal{Q}$ of $\frak{M}$ come from the $d=3$ dualization
of the $n=n_{V}+1$ $d=4$ vector fields. Note that, consistent with (\ref
{HSS-rel}), the $\Omega $-tensor of special quaternionic geometry never
vanishes \cite{Ferrara-Sabharwal}. Moreover, all symmetric non-compact
quaternionic spaces are \textit{special} quaternionic, with the exception of
the \textit{quaternionic projective spaces} \cite{Wolf}
\begin{equation}
\mathbb{HP}^{n}\equiv \frac{USp\left( 2,2n\right) }{USp\left( 2\right)
\times USp\left( 2n\right) },
\end{equation}
which are also the unique example of symmetric quaternionic space with
vanishing $\Omega $-tensor and thus vanishing $\mathcal{L}^{+4}$ (see Sec. 8
of \cite{HSS-2}).

\section*{Acknowledgments}

We would like to thank M. J. Duff, E. Sokatchev, R. Stora, M. Trigiante and
V. S. Varadarajan for enlightening discussions.

A. Y. would like to thank CERN Theory Division for kind hospitality.

The work of S. F. is supported by the ERC Advanced Grant no. 226455, \textit{%
``Supersymmetry, Quantum Gravity and Gauge Fields''} (\textit{SUPERFIELDS}).

The work of A. Y. is supported by the ERC Advanced Grant no. 226455, \textit{%
``Supersymmetry, Quantum Gravity and Gauge Fields''} (\textit{SUPERFIELDS}).

\end{document}